\documentstyle[11pt,newpasp,twoside,epsf]{article}
\def\edcomment#1{\iffalse\marginpar{\raggedright\sl#1\/}\else\relax\fi}
\reversemarginpar

\begin{document}
\title{Photometric Observations of $\omega$ Centauri: Multi-Wavelength
Observations of Evolved Stars}
 \author{Joanne Hughes}
\affil{Everett Community College, Physics Department, 2000 Tower Street,
Everett, WA 98201, U.S.A.}
\author{George Wallerstein}
\affil{University of Washington, Astronomy Department, Box 351580, Seattle,
WA 98195-1580}
\author{Floor van Leeuwen}
\affil{Institute of Astronomy, The Observatories, Madingley Road, Cambridge, CB3 0HA, UK}

\begin{abstract}
We present multi-wavelength observations of the northern population of $\omega$
Cen from the main-sequence turn-off to high on the red giant branch. We show
that the best information about the metallicity and age of the stars can be
gained from combining vby, B-I and V-I colors (in the absence of spectroscopy).
We confirm our results for the main-sequence turn-off stars: there is at least a 3 Gyr age spread,
which may be as large as 5-8 Gyr, as suggested by Hughes \& Wallerstein
(2000) and Hilker \& Richtler (2000). We find that
B-I colors can be affected by excessive CN-content (which can vary at a given
value of [Fe/H]). 
We use proper motion studies to confirm cluster membership  at and
above the level of the horizontal branch, and we show that the age spread is 
maintained amongst stars from the subgiant branch through the red giants. 
We support previous findings that there is another red giant branch, redder
(Pancino et al. 2000), and younger than the main giant branch 
(Hilker \& Richtler 2000) but containing few stars. 
Even though the spatial distribution of the metal-rich stellar
population is different from that of the metal-poor population (which could
be achieved by capturing a smaller metal-rich star cluster, as suggested by
Pancino et al. 2000), it is likely that $\omega$ Cen is the core of a captured dwarf
galaxy. 
\end{abstract}

\section{Introduction}

Why does $\omega$ Cen have a spread in metallicity? Can the largest and most
massive globular cluster in our galaxy (about $3\times 10^6 M_{\sun}$), 
be a composite object, made up of separate star
clusters which merged after they finished forming stars (Norris et al. 1997;
Norris et al. 1996; Jurcsik 1998)?
One or more mergers of star clusters could give an age spread along with 
a metallicity spread, but
this seems dynamically unlikely (Thurl \& Johnston 2000, Thurl 2002).
The structure of $\omega$ Cen is also fairly ``loose'' 
compared to more centrally condensed clusters like 47 Tuc (Mayor et al. 1997).
Bessell \& Norris (1976) suggested that the
gas cloud which gave birth to $\omega$ Cen was chemically inhomogeneous, and Norris
et al. (1997) also discussed the possibility the cluster could have formed when cloud fragments
with different metallicities mixed. All these scenarios could give rise to the 
metallicity spread which has been observed by many researchers, but most notably
detailed by Smith (1987) and Suntzeff \& Kraft (1996). 

Currently, the most popular theory for the origin of $\omega$ Cen involves 
the Milky Way capturing and disrupting a dwarf galaxy (possibly a dwarf 
elliptical like the Sagittarius system), which is dynamically more likely than merging separate
clusters (Majewski  1999; Dinescu et al. 1999; Lee et al. 1999). 
Self-enrichment was considered by Morgan \& Lake (1989);
their theoretical calculations predicted that the ejecta from 330 Type II SN 
must have been retained to increase the metallicity over the range observed.
Norris et al. (1996; 1997) discussed self-enrichment, but preferred the
merger hypothesis.
Hughes \& Wallerstein (2000) found evidence that $\omega$ Cen experienced a period of
self-enrichment which lasted several billion years; their results from
Str\"omgren photometry were confirmed by Hilker \& Richtler (2000), with both
studies concluding that there was an age spread as well a metallicity spread. 
From these studies, it is clear that the more metal-rich stars are several billion years younger than the
most metal-poor stars in $\omega$ Cen. 

Not only does $\omega$ Cen have a spread in [Fe/H], there is also a well-known
variation in the CN-abundances (Smith et al. 2000, and references therein).
Another interesting question arises when we consider if the surfaces of the
stars could have been polluted by material from evolved cluster members
(Ventura et al. 2001).

In this article, we will describe briefly the results from Hughes \& Wallerstein
(2000), which dealt with Str\"omgren photometry of stars at the main-sequence
turn-off (MSTO). We will then turn to the results presented in Hughes, Wallerstein,
\& van Leeuwen (2002) on the MSTO and more evolved stars, including broadband
photometry.

The study carried out by Hughes \& Wallerstein (2000) on the MSTO stars showed
that objects with high metallicities ($-1.2<[Fe/H]<-0.5$) had an average age of
10~Gyr, those in the middle group ($-1.6<[Fe/H]<-1.2$) were about 12~Gyr old, 
and the most metal-poor stars ($-2.2<[Fe/H]<-1.6$) were 12-14~Gyr in age. 
The distance modulus used was $m-M=13.77$ with $E(B-V)=0.15$).
The Str\"omgren data and VandenBerg et al.'s (2000) isochrones showed no evidence
for a merger in the metallicity distribution (i.e., there was no strong bimodality). 

Hughes \& Wallerstein (2000) also
took data of a single-metallicity cluster, NGC~6397, deliberately making the data noisy.
This was achieved by only taking a few 900-second $vby$ exposures of the cluster and not coadding the 
frames. By this method, the distribution of the metallicities was broadened (derived from
the photometry), and we used this data set for comparison rather than
 simulating data. Using the
real data gave a way of forcing a
cluster with no known metallicity variation or age spread to mimic the width
of the distribution of
$\omega$ Cen. Hughes \& Wallerstein (2000) confirmed that NGC~6397 was
a single-metallicity cluster with no discernible age spread, whereas $\omega$
Cen had an average metallicity of $[Fe/H]=-1.5$, a metallicity range of 
$-2.5<[Fe/H]<-0.5$, and an apparent age spread of 3-5~Gyr.

\section{Observations}

We observed one field $\sim 25\arcmin$ north of the cluster center $\sim 90$ times, 
at $vbyBI$ wavelengths, transforming $y$ to $V$. The seeing at CTIO in May 1996 was
$1 - 1\farcs 5$ and the plate scale was $0\farcs 396$ per pixel 
on the
Tek~2048~$\#3$ chip. The field size is 13\farcm 5 $\times $ 13\farcm 5, but the
Str\"omgren filters used at the time were $2\times 2$ inches (instead of $3\times 3$
inches for the broadband filters) and caused vignetting, thus the effective field
of view was
reduced to about 12\farcm 0 $\times$ 12\farcm 0. This field was selected to be
far enough away from the cluster center for the 0.9-m telescope and the DAOPHOT
software to be able to cope with the crowding. In addition, we examined the
IRAS sky flux images at 100\micron\ and 60\micron\ (Wood \& Bates 1993), and
tried to avoid any regions which appeared to have variable extinction.
We took exposures in a field away from the cluster (about $1\deg$ west
of our cluster field, at approximately the same galactic latitude) to
subtract field stars statistically.

\begin{figure}
\plottwo{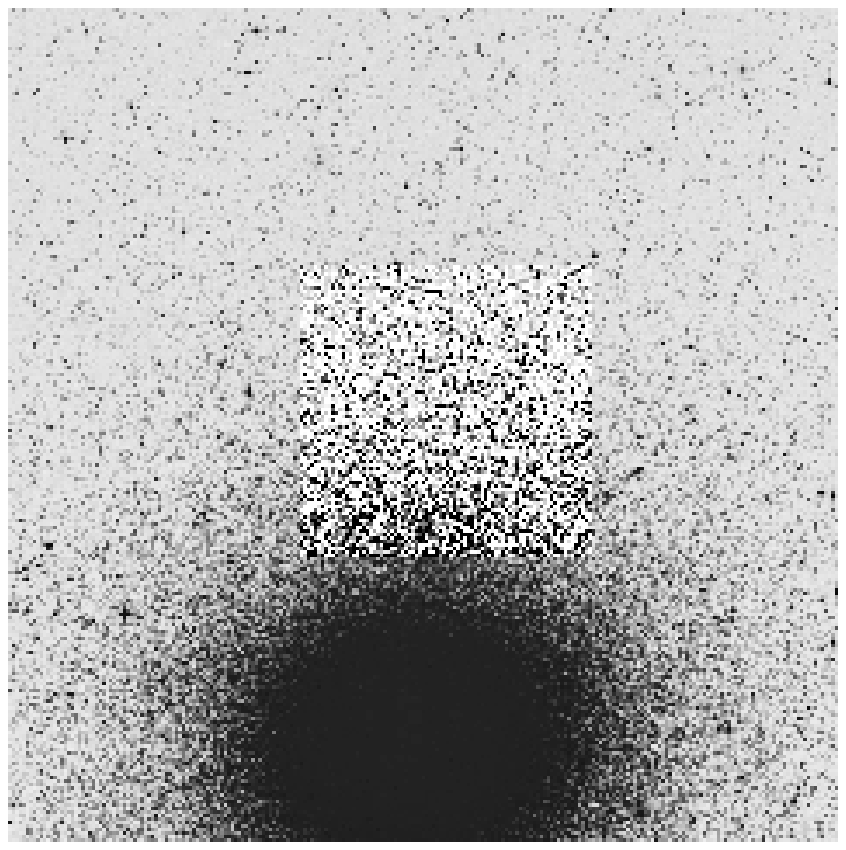}{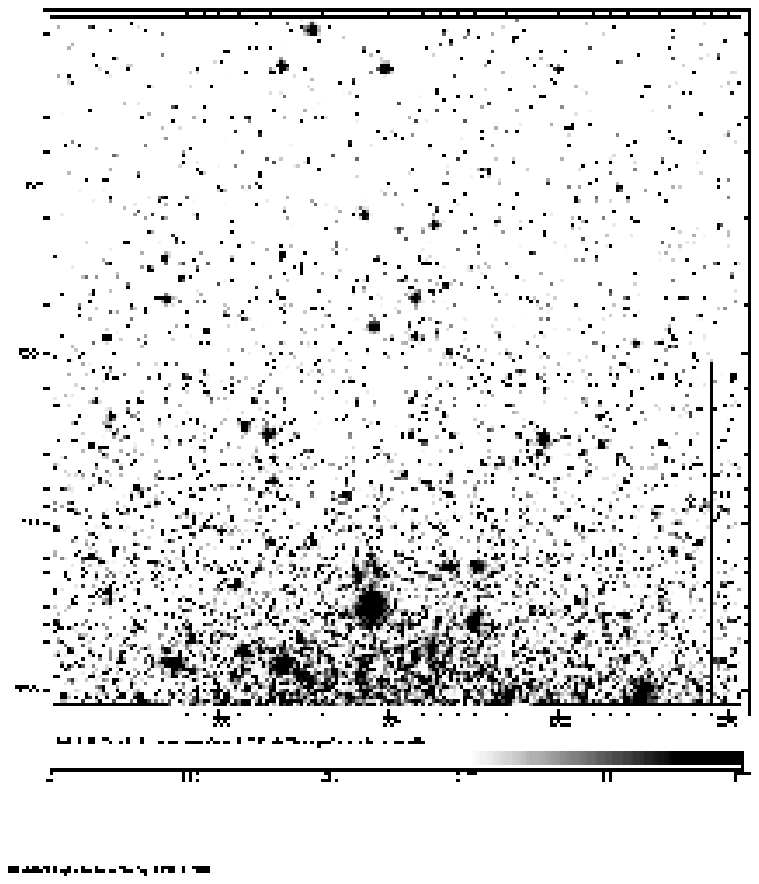}
\caption{\bf a: \rm DSS image of $\omega$ Cen with the 300s I-exposure
identified (area 13.5 square-arcminutes). Here north is up and east is to the
left.
\bf b: \rm  The 300s I-image of the field used in this study is shown with the
pixel scale;
the image has a resolution of  0\farcs 396 per pixel.}
\end{figure}
 
We obtained $vbyBVI$ photometry on 2554 sources in the on-cluster field, and 1739 stars survived
the cleaning process (for details, see Hughes \& Wallerstein 2000). The whole
data set is shown in Figure~2. 

\begin{figure}
\plotone{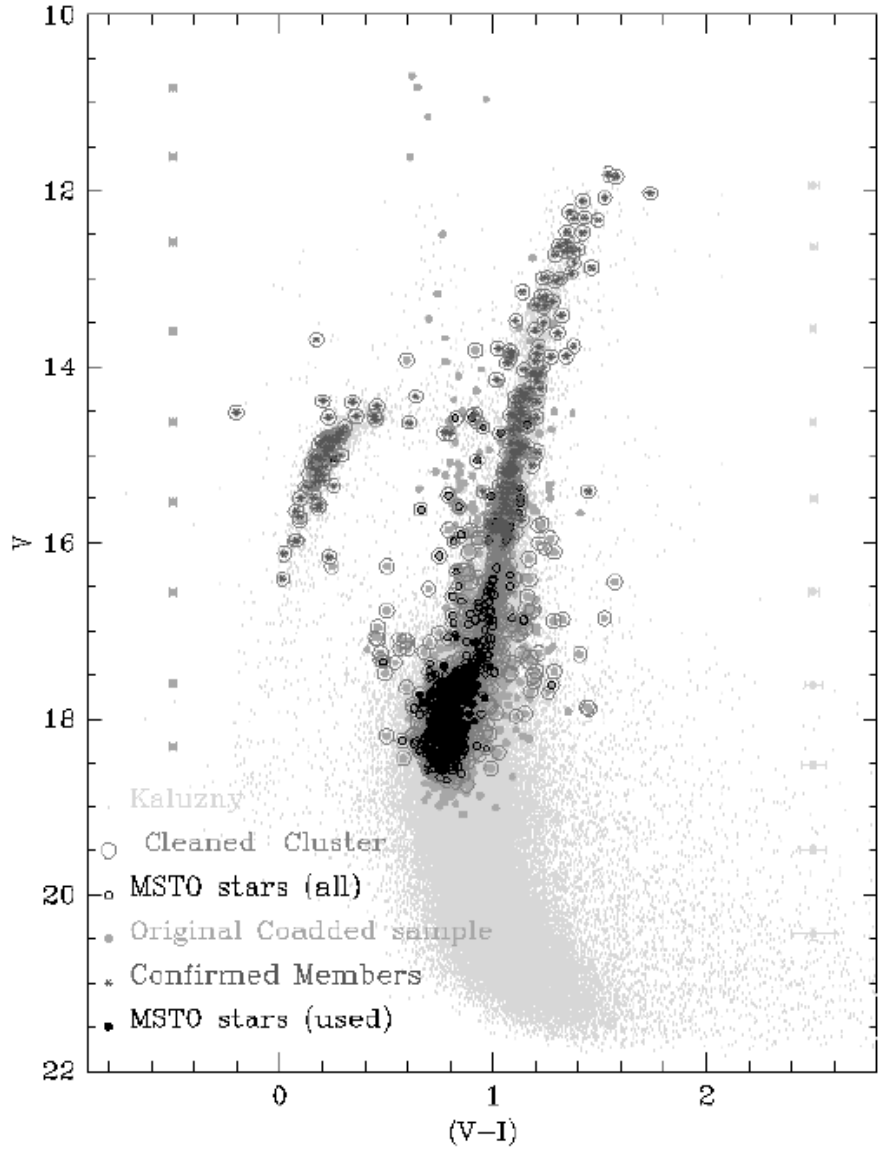}
\caption{Color-magnitude diagram of our sample, along with the data of
Kaluzny et al. (1996; 1997a; 1997b). Our uncertainties are smaller because of the coadding process,
and the fact that we were not trying to get as faint as possible. The aim was to
obtain better than 1.5\%\ photometry at the MSTO and brighter. The confirmed 
cluster
members were identified from the radial velocity and proper motion data of
van Leeuwen et al. (2001).}
\end{figure}

It is almost impossible to recognize an age-abundance correlation among the
giants or RR~Lyrae stars. However, where the isochrones separate and flatten
out at the MSTO, we can see if the more metal-rich stars form later, as
would be expected from self-enrichment (Morgan \& Lake 1989). Figure~3 shows
that if there was a metallicity spread, and no age spread, the more metal-rich
stars should fall to the red side of the color-magnitude diagram (CMD).

\begin{figure}
\plotone{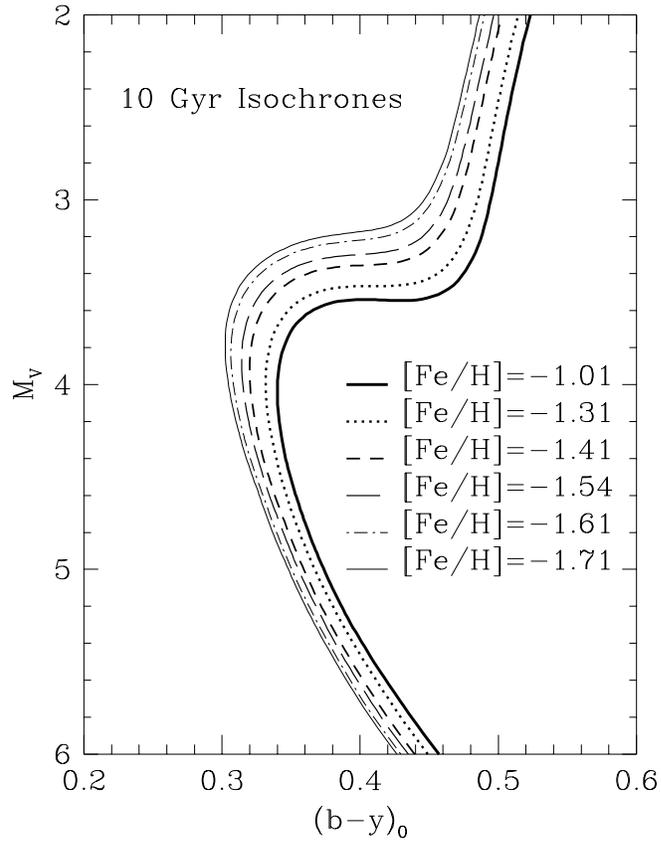}
\caption{Color-magnitude diagram showing the effect of varying the input
metallicity on the color of the main-sequence turn-off. We show the
10 Gyr isochrones for various metallicities from VandenBerg et al. (2000).
If stars are the same age, the redder objects should be more metal-rich.}
\end{figure}

Two colors (or color
differences) should be measured, one that reveals the age and one that
correlates with $[Fe/H]$. One of the photometric systems well designed for
this purpose is the
Str\"omgren system. In principal, the use of the $uvby$ system is
very simple, but in practice, the accuracy of the data is vital to the
success of the experiment. 

Ideally, it would be
desirable to observe in the $u$-band to obtain the $c_1$-index (which
measures surface gravity), but it was
not practical at the 0.9-m telescope because of the transmission
of the filter and the faintness of many of our targets in this region.
The color $(b-y)$ is
sensitive to temperature; $(v-b)$ is sensitive to line-blanketing.
The metallicity index $m_1=(v-b)-(b-y)$ is thus the line-blanketing at a given
$T_{eff}$, and correlates with metallicity. The interstellar reddening is nearly
the same in $(v-b)$ and $(b-y)$, making $m_1$ almost insensitive to interstellar
extinction. 

We define:
\begin{displaymath}
E(m_1)=-0.32E(b-y);
\end{displaymath}
\begin{displaymath}
m_0=m_1+0.32E(b-y);
\end{displaymath}
which is the reddening-free metallicity index, where
\begin{displaymath}
E(b-y)=0.7E(B-V).
\end{displaymath}

In the range of $-0.5 < [Fe/H] -2.0$, the slope of the
$m_1$ vs. $[Fe/H]$ relation is $\Delta [Fe/H] = .056 \Delta m_1$.
Hence, we expect a total range in  $m_1$ of $0.125 \; mag.$
The models of VandenBerg \& Bell (1985), and VandenBerg et al. (2000) show that, at the main-sequence
turn-off, the color-magnitude diagram is nearly vertical. The color index
$(b-y)$ is sensitive to age with a slope given by $\Delta (b-y) = 0.010 \Delta
age(Gyrs)$ between 10 and 15 Gyrs. This is rather small to detect with
sufficient accuracy, but the color $(B-I)$ in the broadband filters
of the BVRI system shows a sensitivity of 0.025 for each Gyr, which is much
easier to recognize. The effect of abundance on $(B-I)$ can be compensated
for from the tables of VandenBerg et al. (2000) once the metallicity is established
from the $m_1$-index. We had thought that the best broadband index to obtain
was $(B-I)$, but it became obvious that CN was affecting the B-magnitudes of
many stars. The advantage of the longer baseline in temperature
is offset by the dependence on chemical composition. 

The Harris B-filter extends from 5000\AA\ to about 3700\AA , hence
many features are included. In particular, many metallic lines are found here,
Ca H and K, and also CN(3883\AA\ and 4216\AA ) and CH(4300\AA ). Suntzeff (1981)
looked at giants in M3 and M13, and found that asymptotic giant branch (AGB)
 stars have C-abundances down
by a factor for 2 compared with the subgiant branch at the same temperature. In M13, almost
all AGB stars and the tip of the red giant branch (RGB), are CN-poor,
which would make them look metal-poor on the $m_0\; vs.\; (b-y)_0$ diagrams.
If CN is weak, the stars will look like they have weak metal lines across the
B- and v-bands. It is for these reasons that we have chosen to use $(V-I)$ as
the temperature index, and have converted $y$ to the V-band (see Hughes \&
Wallerstein 2000, and references therein). 

Before converting photometric indices to metallicities, and finding ages
from the model grids of VandenBerg et al. (2000), we have to find a distance
modulus and extinction for the cluster. Currently, there is a discrepancy
between distances derived from different methods. At one
extreme,  some favor an apparent distance modulus of $m-M=14.10$, $E(B-V)=0.12$ from RRab stars
(Rey at al. 2000); Thompson et al. (2001) find $m-M=14.05$ and
$E(B-V)=0.13$ from model fits to the eclipsing binary, OGLE 17, which gives
an absolute distance modulus of $m-M=13.65$; and Caputo
et al. (2000) find the pulsational distance to $\omega$ Cen from C-type RR Lyrae
variables to be $m-M=14.01 \pm 0.12$. Lub (2002) looked at horizontal branch
(HB) stars in the cluster and determined $E(B-V)=0.11 \pm 0.01$. He also noted
that $HI$ measurements imply $E(B-V)=0.12$, infrared dust measurements from
IRAS and COBE give $E(B-V)=0.14$, and ultraviolet observations from 
Whitney et al. (1998) show $E(B-V)=0.15$. Observations of the star ROA~24
by Gonzalez \& Wallerstein (1994) give $E(B-V)=0.18$: this star is in our field,
but was saturated on the CCD. It is possible that the discrepancy between the
measurements of extinction from the UV, visual and IR means that the
extinction might be anomalous in this direction. At the other end of the distance range, the 
proper motion study of van Leeuwen et al. (2001) give a  reduced 
absolute distance
modulus of $m-M=13.36$.

We compared our $V$ vs. $(V-I)$ CMD to the model grids of VandenBerg et al.
(2000), where $[\alpha/Fe]= +0.3$, and $Y=0.2352$. Isochrones and population
functions can be calculated from these grids (Bergbusch and VandenBerg 2001);
these models do not address gravitational settling or radiative acceleration, 
but they are derived using current physics. From Figure~4a, we see the best-fit
to the data is $m-M=13.57$ and $E(B-V)=0.15$. However, we will calculate ages
based on the other distance moduli to examine the effect on the age-metallicity
relationship.

\begin{figure}
\plotone{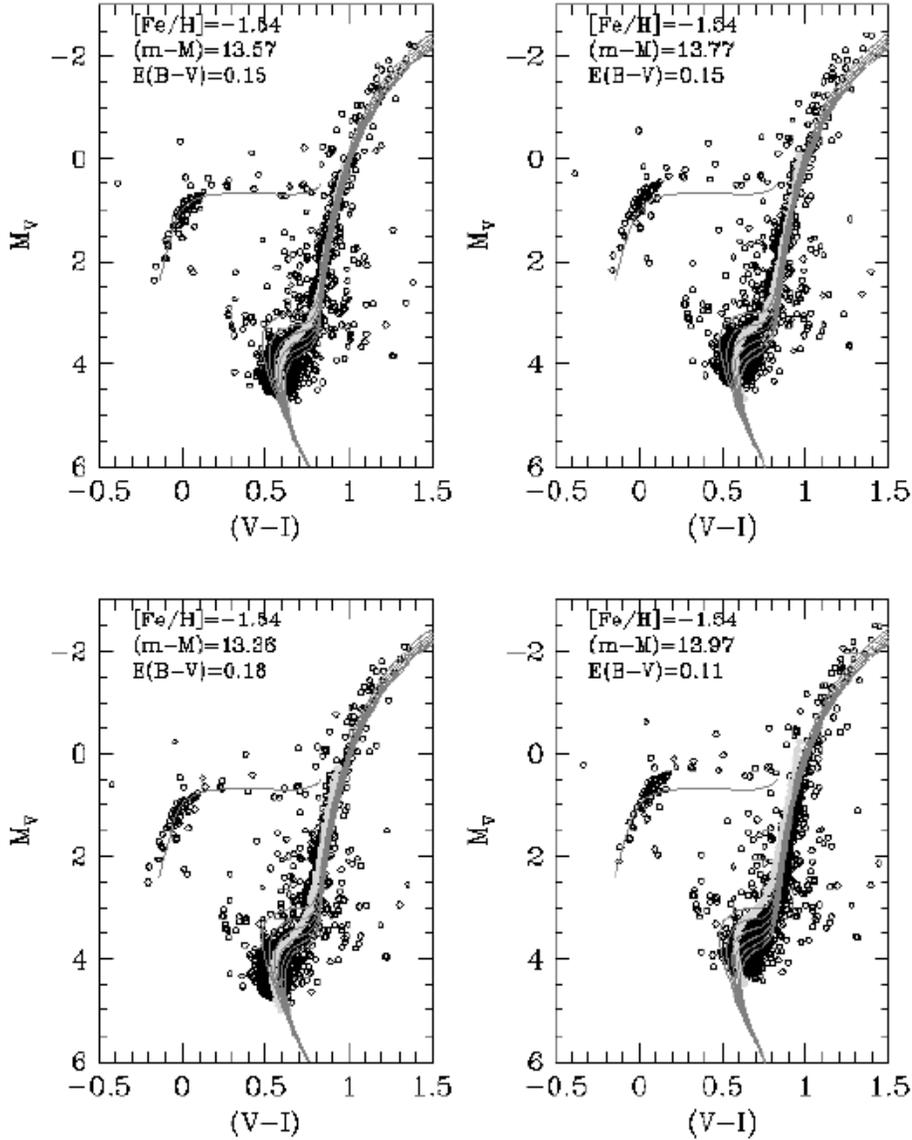}
\caption{Using our ``cleaned'' cluster members, we compare the fiducial 
main-sequence and shape of the horizontal branch to the models of VandenBerg
et al. (2000). The best fit to our data is \bf a: \rm $m-M=13.57$ and 
$E(B-V)=0.15$. Also shown are \bf b: \rm  $m-M=13.77$ with $E(B-V)=0.15$;
\bf c: \rm $m-M=13.36$ with $E(B-V)=0.18$; and \bf d: \rm $m-M=13.97$ with $E(B-V)=0.11$.}
\end{figure}

\section{Metallicity and Age}

Once we have our data set, we must convert the Str\"omgren photometry to 
a metallicity. Mayluto (1994) determined a relationship between $m_1$ and [Fe/H] within the following ranges:
 
\begin{displaymath}
0.22 \leq (b-y) \leq 0.38
\end{displaymath}
\begin{displaymath}
0.03 \leq m_1 \leq 0.22
\end{displaymath}
\begin{displaymath}
0.17 \leq c_1 \leq 0.58
\end{displaymath}
\begin{displaymath}
-3.5 \leq [Fe/H] \leq 0.2
\end{displaymath}
In all these relationships, the authors assume that the $m_1$ and $(b-y)$ 
values are unreddened. 
We define:
\begin{displaymath}
(B-Y)=((b-y)-0.22)/0.16 + 1;
\end{displaymath}
\begin{displaymath}
M_1=(m_1-0.03)/0.19 + 1;
\end{displaymath}
and use
\begin{displaymath}
[Fe/H]=5.7071(B-Y)M_1 - 49.9162(B-Y)logM_1 + 7.9971(B-Y)^2logM_1 
\end{displaymath}
\begin{displaymath}
 - 0.5895(B-Y)^3  - 24.0889(1/M_1)
+ 14.6747
\end{displaymath}
Malyuto (1994) determined an uncertainty of $\sigma_{[Fe/H]}=0.147$.
 
Grebel and Richtler (1992) determined another calibration for giants:
\begin{displaymath}
[Fe/H]={{m_1+a_1(b-y)+a_2}\over{a_3(b-y)+a_4}}
\end{displaymath}
 
where
$a_1=-1.24\pm 0.006, \; a_2=0.294\pm 0.03, \; a_3=0.472\pm 0.04, \;
a_4=-0.118\pm
 0.02$, and these equations again
imply intrinsic (unreddened) colors.
 
Hilker's (2000) calibration for $\omega$ Cen takes the form:
\begin{displaymath}
[Fe/H]_{phot}={{m_0-1.277.(b-y)_0+0.331}\over{0.324.(b-y)_0-0.032}}
\end{displaymath}

\begin{figure}
\plotone{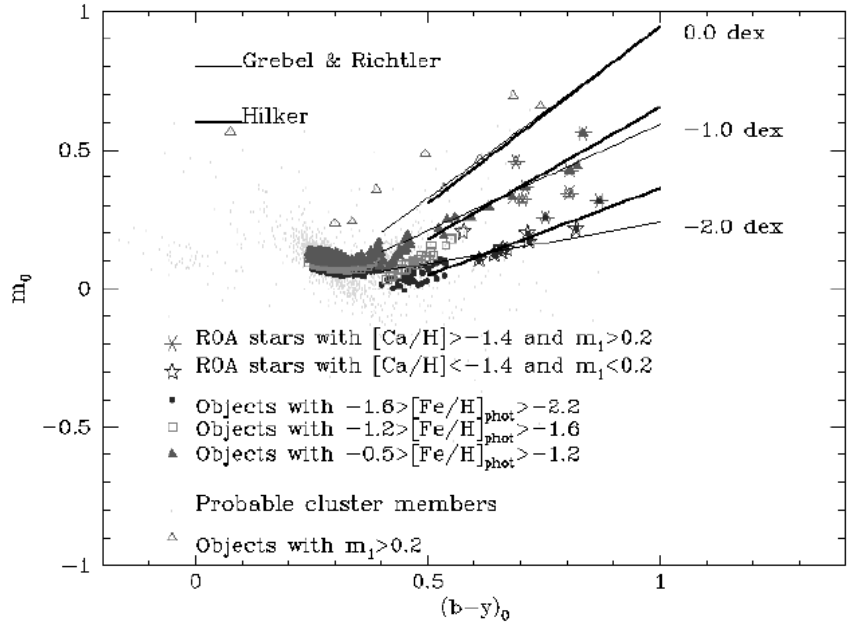}
\caption{Color-color plot of the dereddened metallicity and temperature
indices for $E(B-V)=0.15$. The MSTO stars are selected from the color ranges
specified by Malyuto (1994), and all the proper motion members are classified
by Hilker's (2000) calibration.}
\end{figure}

From Figure~5, we can see that Hilker's (2000) calibration of $[Fe/H]_{phot}$
for the RGB is a better match to all the giant branch stars in $\omega$ Cen than
the Grebel and Richtler (1992) calibration. There is a discontinuity between
the MSTO in their narrow color-range (Malyuto 1994), but improvements will have
to wait for spectra of MSTO stars in $\omega$ Cen. The stars with $m_1>0.2$ which
are proper-motion cluster members are assumed to be part of the ``anomalous''
RGB defined by Pancino et al. (2000). 
The derived values of $[Fe/H]_{phot}$ for MSTO stars are not
very sensitive to extinction.

\begin{figure}
\plotfiddle{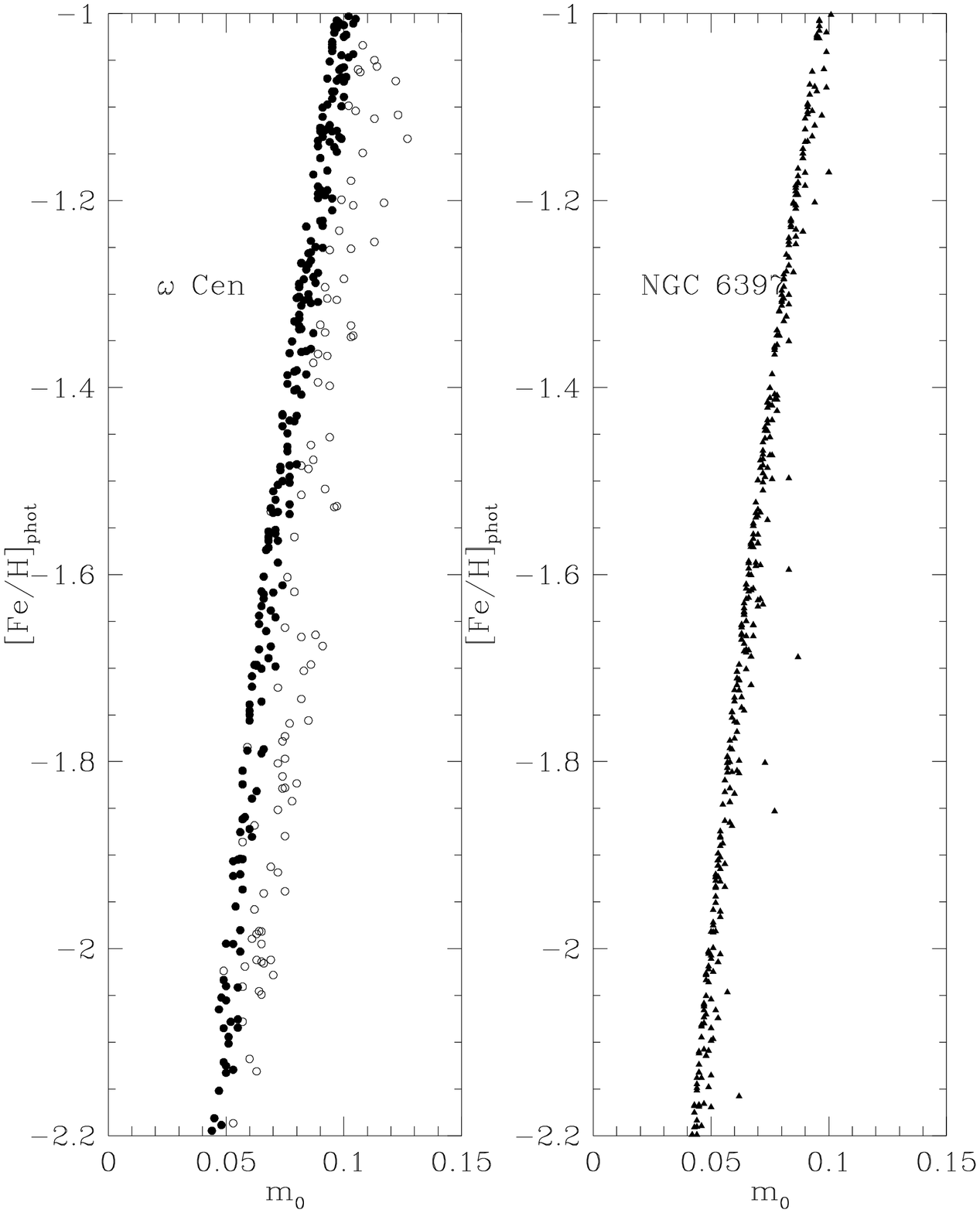}{3in}{0}{50}{35}{-150}{-20}
\caption{\bf a: \rm Plot of $m_0$ vs. $[Fe/H]_{phot}$ for MSTO stars in 
$\omega$ Cen. The filled circles are those stars that fall within the (generous)
noise limits defined by the un-coadded data. By inference, stars which scatter
further from this region (open circles) have done so because of anomalous
chemical composition. \bf b: \rm Plot of $m_0$ vs. $[Fe/H]_{phot}$ for MSTO stars (and
field stars that were not statistically removed) in the ``noisy'' data for
NGC~6397.}
\end{figure}

Figure 6 shows the effect of CN-variations on the data. The open circles
in Figure~6a are likely to be the CN-rich stars. The well-known CN variations
in this cluster (Smith 1987) have also been seen in other clusters such
as M22 (Anthony-Twarog, Twarog \& Craig 1995; Richter, Hilker \& Richtler 1999).
The CN-abundance spread on the RGB is thus shared by the MSTO stars, which implies
the variation is either primordial or the surfaces have been contaminated
by AGB-star ejecta (Ventura et al. 2001). 

As we did for the data set from Hughes \& Wallerstein (2000), we divided the
MSTO population (the CN-normal group) into three groups:
``high'' with $-1.2<[Fe/H]_{phot}<-0.5$; 
``medium'' with $-1.6<[Fe/H]_{phot}<-1.2$); and
``low'' with ($-2.2<[Fe/H]_{phot}<-1.6$). The latter group should be the
first generation of stars formed in $\omega$ Cen. We plot the data set for the
corresponding metallicities for the mean of each group in Figure~7. Most of
the data is fit by the $[\alpha/Fe]=0.3$ models, but there is a suggestion
that the most metal-poor stars are better fit by the $[\alpha/Fe]=0.6$ 
isochrones. The $\alpha$-enhancement seems to be becoming less extreme as
the metallicities approach solar values. The maximum age of the cluster
also fits more comfortably within the cosmological timescales if we use
the $[\alpha/Fe]=0.6$ models for the primordial stars in the cluster.

\begin{figure}
\plotone{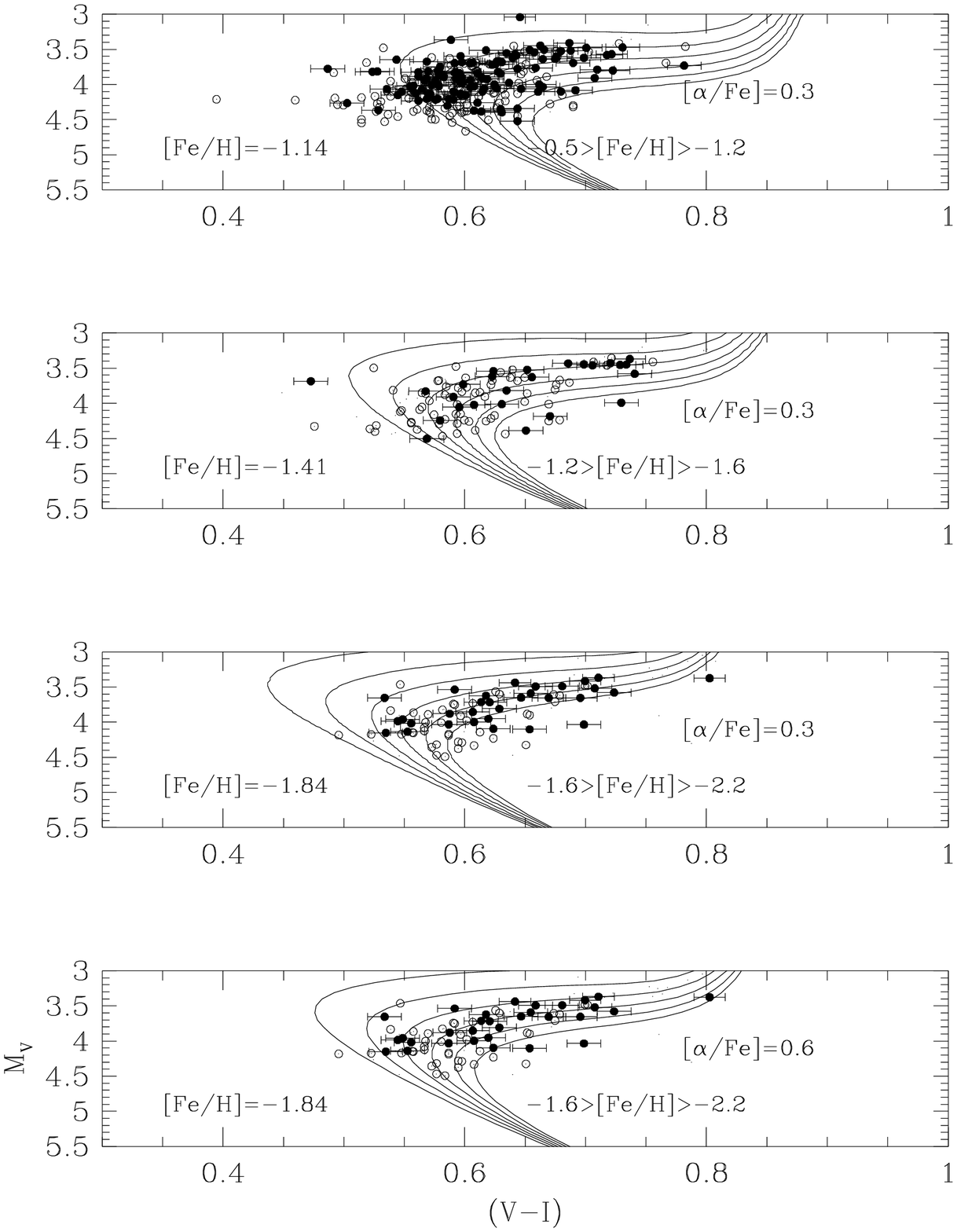}
\caption{The CMDs are shown with the 
best fitting model grids  from VandenBerg et al. (2000). The isochrones
from left to right are 8~Gyr, 10~Gyr, 12~Gyr, 14~Gyr, 16~Gyr, and 18~Gyr.
The distance modulus used is $m-M=13.57$ and the extinction measure is $E(B-V)=0.15$.
\bf a: \rm The high metallicity group has
a weighted mean age of $9.7 \pm 3.0\, Gyr$. \bf b: \rm The medium metallicity
has a weighted mean age of $11.9 \pm 0.9\, Gyr$. \bf c: \rm The metal-poor
stars have a weighted mean age of $15.0 \pm 2.6\, Gyr.$ \bf d: \rm
Using the $[\alpha/Fe]=0.6$ models (Bergbusch and VandenBerg 2001), the
metal-poor stars have a weighted mean age of $12.9 \pm 1.5\, Gyr$.}
\end{figure}

\begin{table}
\caption{Population Ages (in Gyr) for Different Distances and Extinctions}
\begin{tabular}{llllll}
\tableline
$m-M$& $E(B-V)$& High& Medium& 
Low& Low \cr
& & $[\alpha/Fe]=0.3$& $[\alpha/Fe]=0.3$& $[\alpha/Fe]=0.3$& $[\alpha/Fe]=0.6$\cr
\tableline
13.36& 0.18& $9.8 \pm 2.2$& $11.5 \pm 1.1$& $14.7 \pm 2.0$& $13.3 \pm 1.7$\cr
13.57& 0.15& $9.7 \pm 3.0$& $11.9 \pm 0.9$& $15.0 \pm 2.6$& $12.9 \pm 1.5$\cr
13.77& 0.15& $8.9 \pm 2.1$& $10.1 \pm 0.8$& $12.8 \pm 1.6$& $10.9 \pm 1.1$\cr
13.97& 0.11& $9.1 \pm 2.0$& $10.1 \pm 1.0$& $12.8 \pm 1.8$& $11.1 \pm 1.2$\cr
\tableline
\tableline
\end{tabular}
\end{table}

\section{Summary and Conclusions}

Multi-wavelength photometry is necessary in the absence of spectroscopy because
the B-, I-, and v-bands are affected by CN. $(V-I)$ is the best broadband 
temperature index: $(B-I)$ is an indicator of $T_{eff}$ and chemical composition.
Care must be taken with the $m_1$-index because it can be affected by other
sources of opacity than [Fe/H] (Anthony-Twarog et al. 1995; Hughes \&
Wallerstein 2000; Hilker 2000).

We use the RGB to define the metallicity, and the MSTO star to find the age
spread. The first and second generation of stars are separated by at least
2~Gyr, and star formation continued for $\sim 3-5~Gyr$. 
While an initial, non-uniform, composition of the stars in $\omega$ Cen is possible,
it seems more likely that early generations of stars have polluted the cluster
with ejecta from Type II supernovae and AGB stars.

\end{document}